# Probing Berry phase effect in topological surface states

Ya Bai[1,2,*], Yang Jiang[1,2], Wenyang Zheng[1,2], Jiayin Chen[1,2], Shuo Wang[1,2],

Candong Liu[1,2,†], Ruxin Li[1,2,3], Peng Liu[1,2,3,‡]

[1] State Key Laboratory of High Field Laser Physics and CAS Center for Excellence in Ultra-intense Laser Science, Shanghai Institute of Optics and Fine Mechanics, Chinese Academy of Sciences, Shanghai, 201800, China

[2] Center of Materials Science and Optoelectronics Engineering, University of Chinese Academy of Sciences, Beijing, 100049, China.

[3] Zhangjiang Laboratory, Shanghai 201210, China

[*] pipbear@siom.ac.cn; [†]cdliu@siom.ac.cn; [‡]peng@siom.ac.cn

## Abstract

We have observed the Berry phase effect associated with interband coherence in topological surface states (TSSs) using two-color high-harmonic spectroscopy. This Berry phase accumulates along the evolution path of strong field-driven election-hole quasiparticles in electronic bands with strong spin-orbit coupling. By introducing a secondary weak field, we perturb the evolution of Dirac fermions in TSSs and thus provide access to the Berry phase. We observe a significant shift in the oscillation phase of the even-order harmonics from the spectral interferogram. We reveal that such a modulation feature is linked to the geometric phase acquired in the nonperturbative dynamics of TSSs. Furthermore, we show that the overwhelming Berry phase effect can significantly deform the quantum paths of electron-hole pairs, thus enhancing the ability to harness electron spin using lightwaves in quantum materials with strong spin-orbit interactions.

The Berry phase and topology of the Bloch wavefunction, originating from the interplay of internal quantum attributes of crystal electrons [1], hold great significance in classifying novel quantum phases [2,3] and endowing emergent functions [4]. In quantum materials with broken time-reversal or spatial-inversion symmetry, Bloch electrons in cyclic motion necessarily carry a geometric phase [5,6], which defines topological states [2,3,7] and induce various quantum [8] and nonlinear Hall effects [9]. The Berry phase is usually resolved for systems with well-defined Fermi surfaces [7,10] in the absence of diabatic tunnelling between electronic bands. Intense light fields shed new light on nonequilibrium and nonadiabatic dynamics in the geometry and topology of wavefunctions [11-13].

Recent advances in lightwave-driven optical harmonic emission enable the observation of the rapid evolution of Bloch electrons far from equilibrium. Strong lightwaves are capable of evoking rapid evolution of Bloch electrons within a fraction of an optical cycle, imparting information of wavefunctions in transient optical responses such as high-harmonic generation (HHG) and high-order sidebands generation [14-18]. A deeper insight into the electron dynamics on a subcycle time scale can provide an efficient way of manipulating the nonperturbative response of quantum states. When Bloch electrons are exposed to strong laser light, they boost the energy level, gain momentum and encode features of the crystal symmetry and electronic geometry. Hence, the manifested momentum-energy correlation provides in-depth knowledge of the electronic band structure [19,20], Bloch wavefunction [21], Berry curvature [11,12,22-26] and band topology [27,28]. Coherent harmonic emission has revealed a variety of intriguing quantum phenomena, such as the observed quantum coherence and correlation in normal conditions without delicate environmental control. These phenomena include coupling of inter-electronic bands [29,30], quantum path interference [14] and many-body Coulomb interactions [17,31]. So far, despite recent advancements, the endeavor to control subcycle electron dynamics at the quantum level based on geometric properties has been notably challenging. This is due to the intricate nature of the Berry phase embedded in the evolution paths of Bloch electrons, where interference spans across multiple bands.

In this Letter, we aim to probe the Berry phase that involves the coupling between two electronic bands, as visualized in Fig. 1(a). In parametric space, the quantum paths of the electron and hole (*e-h*) quasi-particles between excitation and recombination forms a loop owing to delocalized wavefunctions [32-34]. The geometric phases accumulated along the quantum paths of the *e-h* pairs in their respective bands are compensated by the nontrivial transition dipole phases acquired in their creation and recombination ($\alpha_{\text{ex}}$ and $\alpha_{\text{rc}}$), leading to a gauge-invariant observable,

$$\gamma_{cv} = \int_{\mathcal{C}} d\mathbf{k} \cdot \Delta\mathcal{A}(\mathbf{k}) + \alpha_{rc} - \alpha_{ex}, \tag{2}$$

where $\Delta\mathcal{A}(\mathbf{k}) = \mathcal{A}_c(\mathbf{k}) - \mathcal{A}_v(\mathbf{k})$, $\mathcal{A}_m(k)$ m = c, v represents the Berry connection of the conduction and valence bands, and $\gamma_m = \int_{\mathcal{C}} dk \cdot \mathcal{A}_m(k)$ is the Berry phase along evolution path $\mathcal{C}$. Such a geometric phase is deeply rooted in the dynamical evolution between Bloch states resides in different orbitals, which is a nonadiabatic and noninteger variable. Very recently, it has been observed in transparent dielectric crystal by delicate lightwave control and is referred to as the interband Berry phase [35].

We probe the interband Berry phase in quantum materials with strong interaction between the spin and momentum degrees of freedom of the Bloch electrons. Our idea to interrogating the Berry phase accumulated along the evolution of *e-h* pair is based on two-color high-harmonic spectroscopy of three-dimensional (3D) topological insulators (TIs) [Fig. 1(b)]. In the topological surface states (TSSs) of a 3D TI, the geometric character is embodied in the spin-momentum-locked spin texture governed by the strong spin-orbit coupling (SOC) [36,37]. We show that the observed modulation of even-order harmonics from 3D TI samples can be well reproduced form the transition and evolution of the electron and hole quasiparticles in TSSs, highlighting the crucial role of the interband Berry phase. We find that the relativistic spin-orbit interaction, responsible for shaping the wavefunction geometry [4,7], has a similar magnetic effect that bends the trajectory of electron motion and facilitate the control of electron dynamics via its spin.

To experimentally track the geometric phase encoded in HHG, we utilize a dichromatic pump field composed of a mid-infrared (MIR) field (0.32 eV) and its second harmonic (SH) field (0.64 eV) with precisely tunable phase delay $\phi$, see Fig. S1 in the Supplementary Material (SM) [38]. The SH field can perturb the motion of *e-h* pairs and encode the phase variation in measurable optical emission [Fig. 1(a)]. Through the interaction between the strong laser field and 3D TIs, the produced nonperturbative high-harmonic (HH) emission serves as an observable to detect geometric phase. Figure 2(a) shows the harmonic spectra versus the relative delay of the linearly polarized two-color field (peak field of $F_\omega = 5.2$ MV cm$^{-1}$, $F_{2\omega} = 140$ kV cm$^{-1}$). The HH spectrum is produced from quaternary tetradymite Bi$_{1.5}$Sb$_{0.5}$Te$_{1.7}$Se$_{1.3}$ (BSTS) [41,42] at an orientation angle of 0° ($\bar{\Gamma} - \bar{M}$ direction). Clearly, the even-order harmonics (H4-H8) sensitively depend on the relative phase delay $\phi$ with a modulation period equal to one cycle of the SH field ($T_{2\omega}$), whereas the odd-order harmonics remain unchanged. Note that the modulation of even-order harmonics is almost out-of-phase. In Fig. 2(c) we extracted the oscillation phase $\Phi_{HH}$ of the $T_{2\omega}$ modulation by using the Fourier analysis, as shown. We also found that such a phenomenon exists within a wide range of fundamental field strengths (see Fig. S2 in SM [38]). The significant shift of $\Phi_{HH}$ is rarely observed in inversion-symmetric solids, except in crystals with peculiar types of adjacent bands [29].

In comparison, only a weakly modulated HHG is measured for the crystal angle of $90°$ ($\bar{\Gamma} - \bar{K}$ direction) with a modulation period of $T_{2\omega}/2$ (see Fig. S3 in SM [38]). This is because the mirror symmetry makes the electron dynamics along the $\bar{\Gamma} - \bar{K}$ direction symmetric, and only when a weak SH field is added dose the even-order harmonic with periodic modulation appears. Similar periodic modulations ($T_{2\omega}/2$) have been reported for trivial insulators and semiconductors, which has been attributed to the symmetry properties of electronic states [43,44]. In contrast, for the measurements in the $\bar{\Gamma} - \bar{M}$ direction, the $T_{2\omega}$ modulation is directly related to the lack of inversion symmetry of the 3D TI, and hence should be correlated with the dynamics of surface-confined electrons.

To understand how the dichromatic field drives and affects the HH emission, we investigate the ultrafast dynamics of Dirac fermions in TSSs and massive fermions in bulk states separately by solving the semiconductor Bloch equations (SBEs) [32-34]. Both the surface and bulk states are obtained from the tight-binding Hamiltonian in Ref. [45] and the wavefunction of TSSs is given in SM [38]. By accounting for the geometric phase of the TSSs and the evolution of strong light field created *e-h* pairs, the HHs are periodically modulated as a function of the two-color delay. The key features of the even-order harmonics modulation in our observation [Fig. 2(a)] can be captured by the HHG calculated from the TSSs, with both the oscillating even-order harmonic spectra and the extracted $\Phi_{HH}$ showing well consistency [Figs. 2(b) and 2(c)]. While the barely modulated odd-order harmonics in Fig. 2(a) are do not match with the calculation from TSSs, they can be attributed to the dynamics of trivial electronic states embedded in the crystal bulk (see Fig. S8 in SM [38]).

Unsurprisingly, if we remove the Berry connection and dipole phase in calculating the HHG from TSSs, a distinct difference appears between the simulation and the experimental observation (see Fig.S9 in SM [38]). In the results, both the modulation periods of even-order and odd-order harmonics become $T_{2\omega}/2$, which is similar to the outcome for inversion-symmetric bulk states. Evidently, the geometric phase effect is indispensable in reproducing our experimental observations.

To further confirm the correlation between the oscillation phase $\Phi_{HH}$ and electron dynamics in TSSs, we have measured the two-color HH spectra from other samples, including prototypical TI Bi$_2$Se$_3$ (BS) [7] and indium-doped normal insulator (Bi$_{0.8}$In$_{0.2}$)$_2$Se$_3$ (BIS) [46], as displayed in Figs. 2(d) and 2(e). Figure 2(d) shows the interferogram of HHG from BS. Both the signature of periodic modulation and the shift of $\Phi_{HH}$ as the function of even-order harmonic order are still persisted, with even larger slope of $\Phi_{HH}$ compared with that from BSTS. These common features can also be reproduced by the HHG calculation from TSSs [Fig. 2(f)]. In the modulation of even-order harmonics produced from nontopological BIS, a notable difference appears, as the $\Phi_{HH}$ is not correlated with the increase in harmonic orders [Fig. 2(e)]. Despite the absence of topologically protected surface states in our BIS sample, relatively weak even-order harmonics can still arise from doping-induced lattice symmetry breaking and asymmetric laser fields [23,28]. The highest even-order harmonics from BIS reach up to H6, with intensity more than an order of magnitude smaller than those from TI samples (see Fig. S4 in SM [38]). For samples with strong SOC and TSSs, our observations can show distinct features. The SBE simulations reproduce the experimental observations well, hence unveiling the geometric phase effect in TSS dynamics by its association with the modulation of even-order harmonics.

We primarily focus on interband HHG derived from the *e-h* dynamics because of its predominant contribution to even-order harmonics (see SM [38]). For HHG that involves transitions between electronic bands, asymmetric features of the lightwave and Bloch states are embodied in quantum phases. These phases are in turn imprinted on the HH intensity as

$$I_{\text{even}} \propto \sin^2(S_2^o + \sigma), \tag{2}$$

where $S_2^o(t_{\text{ex}}, t_{\text{rc}}) = \int_{t_{\text{ex}}}^{t_{\text{rc}}} \boldsymbol{F}_\omega(\tau) \cdot \Delta \boldsymbol{\mathcal{A}}^{\boldsymbol{k}} d\tau + \alpha_{\text{rc}} - \alpha_{\text{ex}}$ denotes the Berry phase accumulated along the quantum path of the *e-h* pair driven by the strong fundamental field, which is an equivalent form of Eq. (1); $t_{\text{ex}}$ and $t_{\text{rc}}$ are the excitation and recombination times of the *e-h* pair, respectively; and $\sigma(\phi) \propto F_{2\omega} \cos[\phi - \theta(t_{\text{ex}}, t_{\text{rc}})]$ is the extra perturbed phase as a function of the relative phase delay (see SM [38]). In this picture, the fundamental field $F_\omega$ serves as the primary driving force for the creation, acceleration and recombination of the *e-h* pair, whereas the SH field $F_{2\omega}$ only slightly alters its phase along the evolution path. The perturbed phase reflects the tuning of *e-h* dynamics, which involves the time integration of the band group velocity and Berry connection (see SM [38] for more details). Note that, even a small change in $\sigma(\phi)$ can significantly modify HHG, effectively mimicking the function of an interferometer.

By adjusting both the SH field strength and the relative phase, the shape of the driven field can be tailored. Thus, an increased SH field may affect the *e-h* dynamics and subsequently modify the oscillation phase. To illustrate this, we have measured the dependence of the HH modulation on the weak field strength. Figure 3a shows the intensity oscillation of H6 as a function of $F_{2\omega}$. As the SH field varies in the range of $F_{2\omega} = 49$ to 195 kV cm$^{-1}$, the modulation maxima remain basically unchanged, but the modulation depth deepens. Our simulation results from TSSs also confirm these observations [Fig. 3(b)].

To identify the combined effect of the Berry phase and the perturbed phase, the even-order harmonic intensity described in Eq. (2) can be further decomposed as

$$I_{\text{even}} \propto \sin^2(S_2^o) + \sin(2S_2^o)\sigma(\phi) + \cos^2(S_2^o)\sigma^2(\phi), \qquad (3)$$

where three terms are presented: the modulation-free term ($c_0$), $T_{2\omega}$ modulation term ($a_1, \Phi_{\text{HH}}$), and $T_{2\omega}/2$ modulation term ($a_2, \Psi_{\text{HH}}$) (see SM [38]). Clearly, the three components scale differently on the SH field strength, as $\sigma(\phi) \propto F_{2\omega}$. We extract the intensity scaling and oscillation phases ($\Phi_{\text{HH}}$ and $\Psi_{\text{HH}}$) included in the modulation map of H6 in terms of their dependence on the SH field strength [Figs. 3(c) and 3(d), colored circles are the experimental data, black lines are from SBE calculation]. In Fig. 3(c), both the intensity scaling of the three coefficients ($c_0$ $a_1$ and $a_2$) extracted from our experimental observation and the quantum simulation are all consistent with that predicted by the Eq. (3). In particular, the extracted $a_1$ scales linearly with the SH field, $a_1 \propto F_{2\omega}$, which serves as an observable for the interband Berry phase. These results provide further evidence that the additional weak field operates as an effective perturbation in probing the Berry phase encoded in *e-h* dynamics.

Deeper insight into the Berry phase effect can be provided by the analysis of semiclassical conditions of interband HH emission (see SM [38]). In view of its similarity with the recollision physics in atomic HHG [33,47,48], the exploration of recolliding *e-h* wavepackets is essential for uncovering the Berry phase effect in solids. Figure 4(a) shows the harmonic photon energy computed from semiclassical model for four $k_0$ points [$Q_1$, $Q_2$ and $P_1$, $P_2$ in the inset of Fig. 4(a), with $k_{Q_1} = -k_{Q_2}$ and $k_{P_1} = -k_{P_2}$] superimposed on the subcycle resolved time-frequency emission spectra. We calculated the semiclassical trajectory driven by the fundamental field only, since the SH field only slightly alters the phases accumulated along the quantum paths. These four $k_0$ points are chosen according to the geometric phase factor $\mathcal{D}(k)$, which is large for $Q_1$, $Q_2$, but almost negligible for $P_1$, $P_2$. Interestingly, the recombination events of *e-h* pairs that initiate from the two Q points ($Q_1$, $Q_2$) with large $\mathcal{D}(k)$ only occur for the negative field polarity, while the recombination channel is switched off (harmonic photon energy is relatively small) as the pump field turns positive. Such a feature is absent for the two P points ($P_1$, $P_2$) with negligible $\mathcal{D}(k)$, with the HH emission is symmetric for adjacent subcycle.

For a visual presentation of how the Berry phase governs the evolution of Dirac fermions, we retrieve the semiclassical trajectories of *e-h* pairs in parametric space [|ΔR|-*k* plane, |ΔR| is the separation of *e-h* wavepackets, Figs. 4(b) and 4(c)]. The recombination event of *e-h* pair [circles in Figs. 4(b) and 4(c)] displays a clear signature of imperfect recollision [33]; that is, the emission occurs even if the centres of delocalized *e-h* wavepackets do not overlap, i.e., |ΔR| ≠ 0. In Fig. 4(b), as we remove the geometric phase term ($\mathcal{D}(k) \equiv 0$), the symmetric classical trajectories (gray lines) are shown for *e-h* pairs created at $Q_1$ and $Q_2$, which are subject to energy band dispersion only. Once the $\mathcal{D}(k)$ is taken into accounted, the trajectories are noticeably deformed [red and purple lines in Fig. 4(b)]. For the trajectory of a *e-h* pair starting from $Q_1$ at $t_{ex}$ = −1.6 fs, the *e-h* pair experiences a large spatial shift at the instant of its birth, hence introducing an additional polarization energy in recombination [red circles in Fig. 4(b)]. In contrast, the evolution of the *e-h* pair starting from $Q_2$ at $t_{ex}$ = −8.0 fs shows an opposite behavior, the geometric phase significantly distort the trajectory and suppresses the recombination energy [Fig. 4(b)]. However, as we tune the excitation time to $t_{ex}$ = −1.6 fs (*e-h* pair started from $Q_2$), the geometric phase boosts the recombination energy again through the modification in *e-h* trajectory. This indicates that the evolution path of the *e-h* pair and the recombination photon energy are remarkably shaped by the Berry phase, makes the HH emission prefers to occur at negative light field polarity. Not surprisingly, such a phenomenon is not obvious for trajectories of the two P points ($P_1$ and $P_2$) with minimized geometric phase factors $\mathcal{D}(k)$ [Fig. 4(c)] (see more details in SM [38], Figs. S10 and S11).

For a TI surface equipped with strong SOC and inversion-symmetry breaking, the Berry phase arises from trajectory-dependent spin rotation [10]. After the creation of an *e-h* pair, their spins are always opposite in direction (see Fig. S12 in SM [38]), hence inducing an instantaneous polarization [33]. As the quasiparticles are further accelerated in response the light field, their spin polarizations are continuously rotate, resulting in a phase change via the coupling between light field and Berry connections accumulating over time (see SM [38]). The overall effect is the prominent geometric phase that accumulates along the full evolution path of the *e-h* pair, which has been observed by our two-color HH spectroscopy.

In conclusion, we present the observation of the Berry phase effect in the TSSs of 3D TIs and provide a microscopic perspective of the lightwave-driven intrinsic degrees of freedom of the Bloch electrons. The interplay of SOC and the lack of inversion symmetry at the TI surface leads to a remarkable Berry phase effect in the subcycle electron dynamics. The reported mechanism enables lightwave manipulation of trajectory-dependent spin evolution, in which the specific trajectory can be switched on and off on a subcycle time scale, and such a phenomenon should also be expected in Rashba–split states and magnetic materials with chiral spin textures. Predictably, the interband Berry phase and relevant quantities are supposed to play crucial roles in exotic phases of matter that have been driven far from its equilibrium, serving as an extra control knob in Floquet-engineered topological phases [49,50] and the transient Hall effect [51]. Exploring the Berry phase effect in quantum many-body systems and non-Abelian Bloch states may further broaden the emergent functions of quantum materials, and promote the potential of harnessing the internal degree of freedom of Bloch electrons using lightwaves.

**Acknowledgements**

This work was supported by the National Natural Science Foundation of China (Grant No. 12174412, 12174413), the Youth Innovation Promotion Association of the Chinese Academy of Sciences (No. 2021241) and the CAS Project for Young Scientists in Basic Research (No. YSBR-091).

**Figures:**

**Figure 1.**

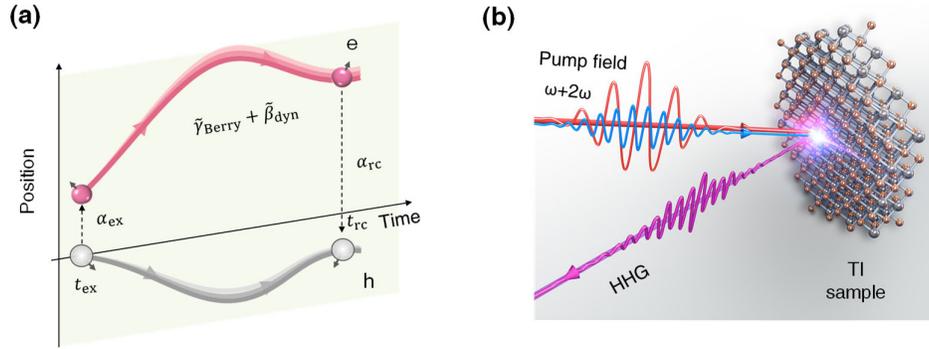

**FIG. 1. Concept of the evolution phases and the experimental configuration.** (a) Illustration of the quantum paths (pink and gray curves) and the phase accumulated in the light field-driven *e-h* pairs in parametric space. The Berry phase $\gamma_{\text{Berry}}$ is acquired along with the rotation of the spin vector, while the dynamic phase $\beta_{\text{dyn}}$ is associated with time-dependent changes in the energy eigenvalues, $\mathcal{E}_m(\boldsymbol{k})$. The diabatic phases $\alpha_{\text{ex}}$ and $\alpha_{\text{rc}}$ arises due to the interband transition at the instant of excitation $t_{\text{ex}}$ and recombination $t_{\text{rc}}$, respectively. The weak SH field perturbs the *e-h* dynamics by tuning the phases to $\tilde{\gamma}_{\text{Berry}}$ and $\tilde{\beta}_{\text{dyn}}$, as a result of subtle changes in detailed evolution paths (pink and gray gradients). (b) Schematic of two-color high-harmonic spectroscopy setup. Two-color field (red and blue curves) with subcycle tunable delay irradiates the TI surface and produces HHG (purple curve) in the reflection direction.

**Figure 2.**

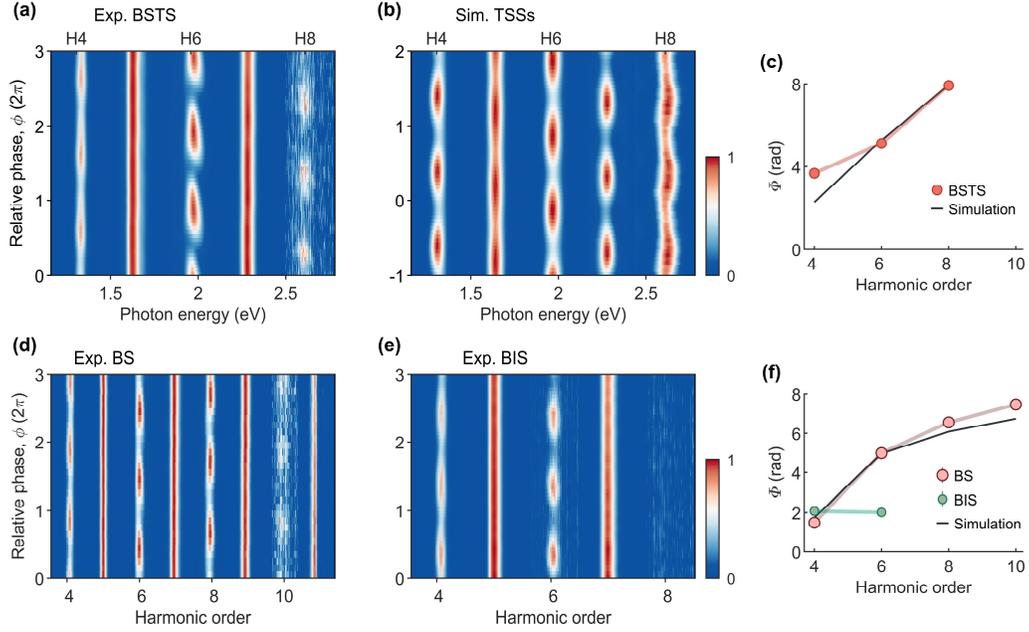

**FIG. 2. Distinguishing TSSs originated contributions from dichromatic HH interferometry.** (a) HHG interferogram from the BSTS surface as a function of time delay between the two-color fields (peak amplitude, $F_\omega$ = 5.2 MV cm$^{-1}$, $F_{2\omega}$ = 140 kV cm$^{-1}$). Each harmonic order is normalized separately. The spectrum image of H8 shows noticeable granular noise due to its low intensity. (b) Simulated HH spectra from light driven TSSs as a function of two-color delay at pump field of $F_\omega$ = 5.5 MV cm$^{-1}$ and $F_{2\omega}$ = 120 kV cm$^{-1}$. (c) Comparison of extracted modulation phase from panel (a) (red circles) and the model calculation in (b) (black line). The broken lines are a guide to the eye. (d)-(e) Dependence of HH spectra on SH delay measured from Bi$_2$Se$_3$ (d), and BIS (e). In (d) different SH field strengths are applied ($F_{2\omega}$ = 60 kV cm$^{-1}$, BS) with the same fundamental field ($F_\omega$ = 5.2 MV cm$^{-1}$). In (e) the pump field strengths are $F_\omega$ = 5.2 MV cm$^{-1}$, $F_{2\omega}$ = 98 kV cm$^{-1}$. (f) Extracted modulation phases from (d) (pink circles, BS) and (e) (green circles, BIS), together with that from SBE simulation (black line). To better match the results in (a) and (d), the spin-flip parameter A$_{12}$ and the intralayer hopping parameter A$_{14}$ take different values in tight-binding Hamiltonian. In the simulation of HHG from the BSTS, A$_{12}$ = 0.196 eV and A$_{14}$ = 0.0482 eV, while for that from the BS, A$_{12}$ = 0.224 eV and A$_{14}$ = 0.0551 eV (see SM [38]).

**Figure 3.**

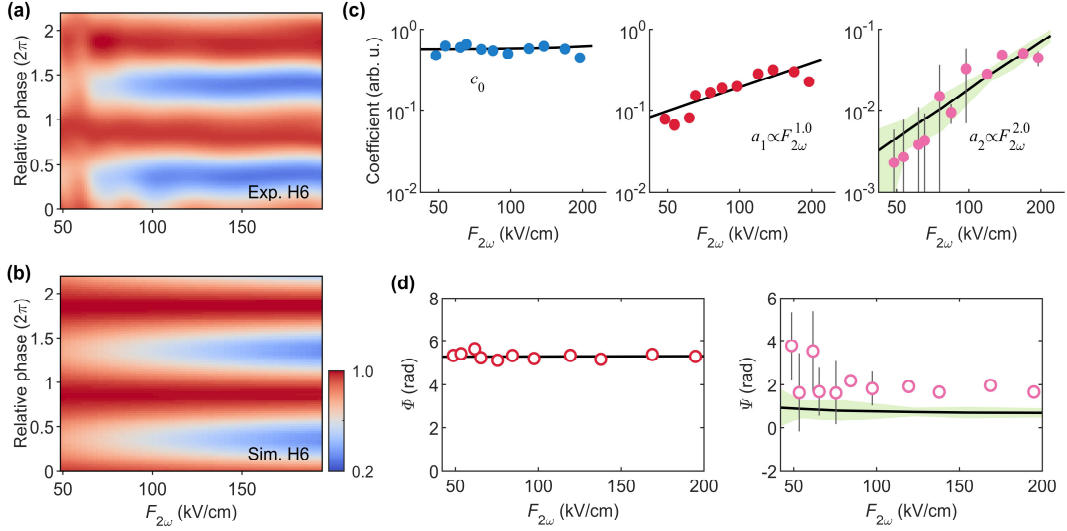

**FIG. 3. Disentangling of the Berry phase effect in HH oscillation.** (a)-(b) The intensity modulation of H6 versus the two-color field delay, at various SH field strengths ranging from 49 to 195 kV cm$^{-1}$. The results of experimental measurement (a) are directly compared with the SBE calculation in TSSs (b). (c) Fourier parameters extracted from the intensity modulation in (a). Three subplots show the fitting coefficients $c_0$ (blue circles), $a_1$ (red circles) and $a_2$ (pink circles) from the experiments compared with those from the simulation (black line). (d) The dependence of the oscillation phases on the SH field. The phases $\Phi$ and $\Psi$ correspond to two different periodic modulations, $T_{2\omega}$ and $T_{2\omega}/2$, respectively. Error bars and green shaded area indicate the standard deviation.

**Figure 4.**

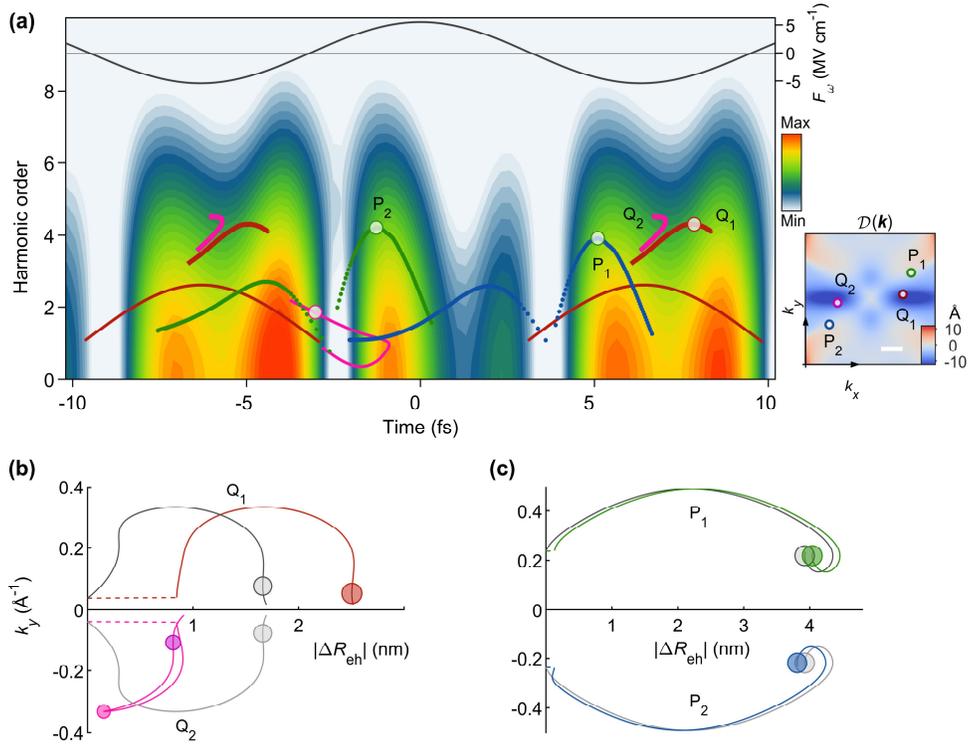

**FIG. 4. Shaping of *e-h* trajectories by the Berry phase effect.** (a) Time-frequency analysis of HHG in the presence of the Berry phase effect. The emission spectrogram is computed by using the Gabor transform with the time window of 1 fs (colored map). Semiclassical harmonic photon energies are extracted from the solution of the saddle-point conditions. The coordinates of the four $\boldsymbol{k}_0 = (k_x, k_y)$ points are $k_{Q_1} = (0.30, 0.038)$, $k_{Q_2} = (-0.30, -0.038)$ and $k_{P_1} = (0.38, 0.24)$, $k_{P_2} = (-0.38, -0.24)$. The black line denotes the driving field $F_\omega$. Inset: the geometric phase factor $\mathcal{D}(\boldsymbol{k})$ of the TSSs in the 2D Brillouin zone, with the four $\boldsymbol{k}_0$ points of interest are marked as dots. A white scale bar at the bottom is 0.2 Å$^{-1}$. (b),(c) Semiclassical trajectories in parametric space, illustrating the time-dependent crystal momentum $k_y$ as a function of the relative separation $|\Delta R|$ of *e-h* pair. Calculated trajectories initiated in Q$_1$ and Q$_2$ (b) excited at $t_{ex} = -1.6$ fs (red) and $-8.0$ fs (purple), and in P$_1$ and P$_2$ (c) excited at $t_{ex} = -7.4$ fs (green) and $-13.8$ fs (blue), respectively, are displayed subjected to the fundamental driver of 5.5 MV cm$^{-1}$. The gray lines in (b) and (c) display the trajectories with $\mathcal{D}(\boldsymbol{k})$ removed. The circles represent the interband HH emission event along each trajectory. The coordinate of each circle marks the $k_y(t_{rc})$ and the $|\Delta R(t_{rc})|$ at the instant of recombination, with the diameter representing the emitted harmonic photon energy. The recombination of the *e-h* pair usually occurs more than once within an optical cycle [see the two purple circles in panel (d)].